# The ICO Phenomenon and Its Relationships with Ethereum Smart Contract Environment

Gianni Fenu, Lodovica Marchesi, Michele Marchesi and Roberto Tonelli
Dept. of Mathematics and Computer Science
University of Cagliari
Cagliari, Italy
{fenu@, l.marchesi@studenti., marchesi@, roberto.tonelli@dsf.}unica.it

*Abstract*—Initial Coin Offerings (ICO) are public offers of new cryptocurrencies in exchange of existing ones, aimed to finance projects in the blockchain development arena. In the last 8 months of 2017, the total amount gathered by ICOs exceeded 4 billion US$, and overcame the venture capital funnelled toward high tech initiatives in the same period. A high percentage of ICOS is managed through Smart Contracts running on Ethereum blockchain, and in particular to ERC-20 Token Standard Contract. In this work we examine 1388 ICOs, published on December 31, 2017 on icobench.com Web site, gathering information relevant to the assessment of their quality and software development management, including data on their development teams. We also study, at the same date, the financial data of 450 ICO tokens available on coinmarketcap.com Web site, among which 355 tokens are managed on Ethereum blochain. We define success criteria for the ICOs, based on the funds actually gathered, and on the behavior of the price of the related tokens, finding the factors that most likely influence the ICO success likeliness.

## I. INTRODUCTION

Recently, the cryptocurrencies phenomenon has become widespread, in terms of adoption, number of available currencies and market capitalization. Made possible by blockchain technology, which ensures trust, security, pseudo-anonimity and immutability through strong cryptography and a decentralized, peer-to-peer approach, a cryptocurrency can be easily dispatched from the initial owner to another person, in whatever part of the world, in matter of minutes and with no intermediary whose behavior can also be modeled using a Petri Net approach [1].. These features make cryptocurrencies ideal also for crowfunding purposes, leading to the so called ICO phenomenon.

Initial Coin Offerings (ICO) are public offers of new cryptocurrencies in exchange of existing ones, aimed to finance projects, mostly in the blockchain development arena. Despite being totally unregulated, and even banned in several countries, the easiness of sending funds through blockchain transactions, and the hope to get very high returns even before the business initiative reaches the market – because ICO tokens are traded immediately on cryptocurrency exchanges – made the ICO phenomenon explode. In the last 8 months of 2017, the total amount raised by ICOs exceeded 4 billion US$, and overcame the venture capital funneled toward high tech initiatives in the same period [1]. ICO are usually characterized by the following features: a business idea, typically explained in a white paper, a proposer team, a target sum to be collected, a given number of "tokens", that is a new cryptocurrency, to be given to subscribers according to a predetermined exchange rate with one or more existing cryptocurrencies.

Nowadays, a high percentage of ICOS is managed through Smart Contracts running on Ethereum blockchain, and in particular through ERC-20 Token Standard Contract. Cloning an ERC-20 contract, it is very easy to create a new token, issue a given number of tokens, and trade these tokens with Ethers – the Ethereum cryptocurrency, which has a monetary value – according to a given exchange rate. The contract stores the addresses of the token owners, together with the amount of owned tokens, and allows transfers only if the sender shows the ownership of the private key associated to the address.

Since the ICO phenomenon had a boom starting from May-June 2017, only a few research reports, and almost no paper published on scientific journals, has appeared on the subject so far. We can just quote the working papers by Zetzsche et al. [2], and by Adhami et al. [3], that report analyses of ICO features. The former paper is focused on legal and financial risk aspects of ICOs, but its second section contains a taxonomy, and some data about ICOs that the authors claim are continuously updated. In the latter paper 253 ICOs are analyzed, starting from 2014 to August 2017, and the significance of some factors that influence the success of an ICO is studied. Recently, Subramanian [3] quoted the ICOs as an example of decentralized blockchain-based electronic marketplace. The main source of information about blockchains, tokens and ICOs is obviously the Web. Here we can find sites enabling to explore the various blockchains associated to the main cryptocurrencies, including Ethereum's one. We can also find Web sites giving extensive financial information on prices of all the main cryptocurrencies and tokens, and sites specialized in listing the existing ICOs and giving information about them. Often, these sites also evaluate the soundness and likeliness of success of the listed ICOs. One of the most popular among these sites is icobench.com, which evalutes all the listed ICOs, and provides an API to automatically gather information on them.

In this work we examine 1388 ICOs, published on December 31, 2017 on icobench.com Web site, gathering information relevant to the assessment of their quality and software development management [11] [12], including data on their development teams [13]. We also studied, at the same date, the financial data of 450 ICO tokens available on coinmarketcap.com Web site, among which 355 tokens are managed on Ethereum blockhain. We defined success criteria for the ICOs, based on the funds actually gathered, and on the behavior of the price of the related tokens, and studied the factors that most likely influence the ICO success likelihood.

We analyzed some key features of the ICOs, like their country, the kind of business they address, the team size, the ratings obtained by icobench.com site. We found that more than 1000 ICOs are managed on the Ethereum blockchain, mainly following ERC-20 standard. This causes a considerable stress on Ethereum blockchain, confirmed by the analysis of token transactions using the data gathered from ethplorer.io site. The total number of transfer transaction is above 16 million, and the total number of token holders is about 5.5 million. After performing a multivariate analysis of the factors influencing the success, we also found that the ratings of icobench.com site have a high probability to predict the success, as well as some of the countries of origin and the platform.

In the followings, Section II presents the methods used to gather the data, to evaluate ICO success and to analyze the data. Section III presents and discusses the results. Section IV concludes the paper.

## II. METHOD

To perform a massive study of the ICOs characteristics and success factors, we need to gather ICO data from the Web, to establish what data are to be analyzed and how ICO success can be defined, and to analyze the data to draw facts about the factors that determine success. The following subsections give insight on these steps.

### A. Retrieving ICO data from the Internet

The main sources of data we used are of three kinds:

- data about the ICOs themselves, collected by icobench.com site;
- financial data about the ICO tokens traded on main cryptocurrency exchanges, collected by coinmarketcap.com Web site; these data include the address of the token contract on Ethereum blockchain;
- data about token transactions and holders directly collected from Ethereum blockchain using a blockchain explorer (ethplorer.io).

ICO data were massively collected from icobench.com, which kindly granted us permission to access their API calls. icobench.com is one of the main sites giving information about ICOs. As its name suggests, icobench.com also performs an analysis of each submitted ICO, giving both the automated rating of its "Benchy" ICO analyzer robot, and possibly the rating of a pool of experts.

Each ICO shown in the site is provided of a unique integer progressive identifier. We performed an API query for all of these numbers, gathering the whole icobench.com database, in json format. The ICO data include name, token symbol, description, rating, country, start and end dates of the crowfunding, financial data such as the total number of issued tokens and the percentage that is sold in the offer, initial price of the token, platform used, hard and soft cap (maximum and minimum number of tokens to sell), raised money (in US$) if the ICO has finished, data on the team proposing the ICO, main milestones and category. Some of these data, such as short and long description, and milestones are textual descriptions. Others are categorical variables, such as the country, the platform, the category (which can assume many values), and variables related to the team members (name, role, group). The remaining variables are numeric, with different degrees of discretization. Unfortunately, not all ICOs record all variables, so there are several missing data.

Financial data were collected from coinmarketcap.com Web site, which is one of the most popular sites giving almost real-time data on the quotation of the various cryptocurrencies in the world exchanges – an exchange is a Web site where it is possible to buy and sell cryptocurrencies against each others, and against standard currencies. It also has a specific "token" section giving information about the tokens (usually ICO tokens). This information included the address of the token contract in the related blockchain (usually Ethereum).

We gathered the needed information from this Web site using the Python scraping library "Beautiful Soup" [4]. For each listed token we recovered name, symbol, number of tokens, capitalization, Ethereum address (if it is a token managed on Ethereum blockchain), price series (daily closing price in US$, volume and market cap) in a given time interval.

Using the Ethereum address, when present, we query ethplorer.io publicly available APIs, gathering information about the total token supply, the number of token transfers, the number of token holders. Using this Web site, it is also possible to obtain information on each transaction, and each holder, but this is beyond the scope of this paper.

All the recovered data were stored in a database, linking data coming from different sources (icobench.com and coinmarketcap.com) through the name and symbol of the tokens. In some cases, name and/or symbol differ, so it was needed a more sophisticated matching procedure, using the Levenshtein distance [5] between names in the case the symbol is the same. If this distance is below 0.6, we assume a match. Anyway, we checked by hand these matches, and also other possible matches (same name, different symbol; same symbol, names with Levenshtein distance greater than 0.6).

### B. Defining ICO success

Since half of the year 2017, the number of ICOs launched on the market skyrocketed. However, only a fraction of them was able to gather an amount of money according to the needs and hopes of their proposers. Moreover, all successful ICOs after the end of the sale are quoted on some exchange, where they are traded against other currencies, usually Bitcoin. Quite often, in the past, the actual price of the tokens increases a lot after their quotation on the exchange, and this is one of the main attraction factors of ICOs – the fact that the token can be sold with a profit very soon, long before the realization of the business initiative behind the ICO.

In our analysis, we used a dichotomous variable to describe the ICO's success: successful or failed. This is the same approach of the paper by Adhami et al. [2]. They define an ICO as "successful" if it reaches at least the soft cap declared by its proposers. We decided to extend this definition because on one hand our data may lack the value of an ICO's soft cap, and on the other hand several ICOs include provisions allowing to go ahead with the ICO even in the case the soft cap is not reached, and this happens in many cases. Several ICOs are not eligible to be considered, typically because they lack data, or are still in progress. To assess whether an ICO is successful, the criteria we use are the following:

1. we regards as failed an ICO that raised less than 80.000 US$; we regard as undecided – and did not consider in the analysis – an ICO that raised between 80.000 and 200.000 US$; ICOs raising more than 200.000 US$ are considered successful, except in the case they fall in criterion 3.
2. we do not consider ICOs ending in 2018, except the few ones that raised money in 2017 and were stopped; we do not consider ICOs that raised no money and that have no end date;
3. for ICOs with a token provided of a price series long enough in 2017 (at least prices in the whole month of December 2017), we considered as failed the ICOs with a market cap diminished by more than 75% since the beginning of their quotation; the market caps are computed as a moving average of 20 consecutive days, to filter out daily variations.

*C. Analysis of the factors influencing the success*

Among the data associated to an ICO, we chose some factors that could possibly influence its success. These factors are the ratings obtained by icobench.com Web site, the country of origin, the team's size, the opening and closing date, number of tokens sold, the platform, the category and others.

To analyze our dataset we resort to multivariate statistical analysis for dicotomic dependent variables. In fact our target is to measure if and to which extent the collected variables contribute to the success or failure of an ICO prject. Given the dicotomic nature of the target variable, success or not, simple regression analysis and predictive models cannot be applied directly.

We set to one the dependent variable in cases where the ICO has been successful and to zero otherwise. Success or failure can be so tested against the set of independent variables which is the set of variables we collected from icobench.com and the contribution of each variable to success or failure can be evaluated and compared with other variables.

The best suited model is the Logit model, where the logarithm of the odd ratio among success and failure is modeled through a multivariate linear analysis as a linear combination of the independent variables of interest. The model outputs the best fitting coefficients as well as the statistical significance of each variable with respect to ICO's success or failure.

In order to simplify our analysis we filtered the raw data for some variables and concentrated the analysis on part of them. Specifically, for the multivariate analysis we didn't consider the raised founds, which has been already chose as the discriminant variable for the success, the token which is simply a label, the type, whose values are mostly missing or mostly equals, and all other variables with many missing values.

According to the Logit model we define:

$$\ln \frac{P}{P-1} = a + \sum_{i=1}^{N} \beta_i x_i \qquad (1)$$

the Logit model where P represents the success probability, 1-P the failure probability, P/(1-P) the *odd ratio*, and the sum is a linear combination of all independent variables in the vector **x** with coefficients in the vector **beta**. We implemented the computation using the R packages 'lme4' and 'rms', using the general regression model provided by the 'lrm' function.

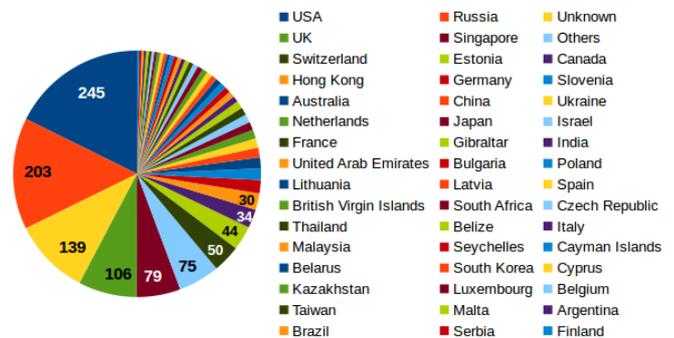

*Fig. 2: The countries of origin of the ICOs.*

At first we targeted the level of significance of independent variables with respect to influencing ICOs success or failure and how much the single coefficients variation can affect the odd ratio against failure.

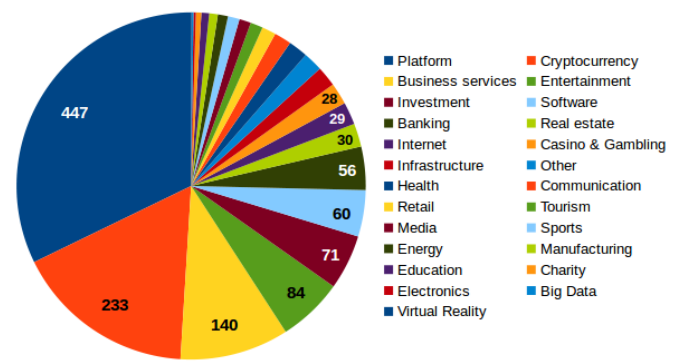

*Fig. 1: The main categories of the ICOs.*

## III. RESULTS AND DISCUSSION

We gathered all ICOs listed on icobench.com Web site on 31/12/2017. Overall, they are 1387. We also gathered information on 450 tokens listed on coinmarketcap.com, and on 355 tokens managed on the Ethereum blockchain and whose data are reported on ethplorer.io site. On these data we

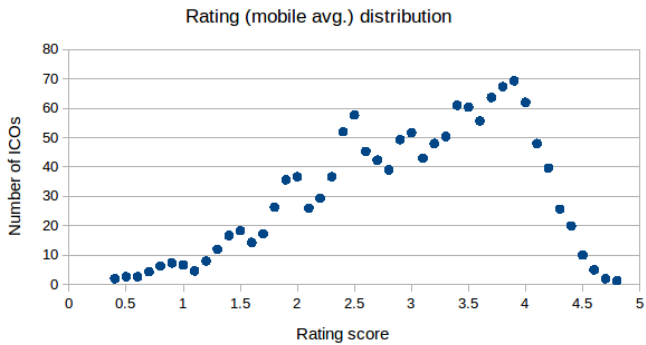

Fig. 3: The distribution of ICO ratings (mobile average of 3 rating scores).

### A. Descriptive statistics of the ICO data

We report in Fig. 1 the countries of origin of the ICOs. As you can see, USA and Russia Federation are the most active countries in proposing ICOs. The country of 139 ICOs is not declared in icobench.com; the countries with less than 4 ICOs are cumulated under the "Others" tag. We note that some relatively small countries, like Singapore, Switzerland, Estonia and Slovenia are very active in proposing ICOs.

Fig. 2 shows the main business category of the ICOs analyzed. Note that icobench.com allows to assign more than one category to an ICO. Here we report just the first category, which we assume is the most expressive of the ICO business target. Most ICOs declares themselves as "platforms" to perform decentralized business. 233 ICOs are new kinds of cryptocurrencies, whereas the remaining categories cover almost all business sectors.

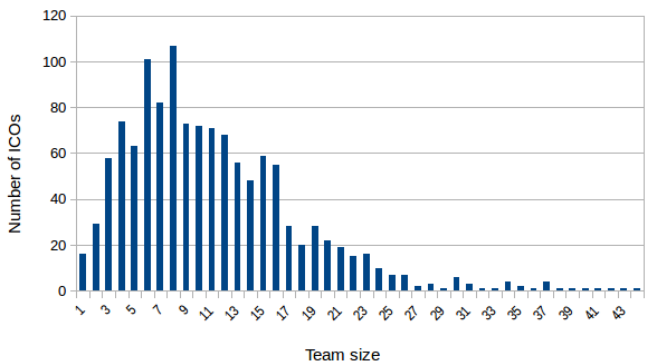

Fig. 4: The team sizes of the studied ICOs.

The distribution of the overall ratings given to the various ICOs is reported in Fig. 3. All considered ICOs have a rating, in most cases given by the robot of icobench.com site. The ratings span between 0 and 5. In the figure we report the centered moving average of 3 rating values, to filter out the noise. As you can see, the distribution is quite regular, with a steady climbing of the rating from the value of about 1.2, until the peak at the value of 3.9; then a steep descent follows, with very few ratings equal or above to 4.5.

In Fig. 4 we report the team size distribution, as declared by ICOs proposers. Note that, in this analysis, we consider the overall team, including business people and advisors, and not only the software development team. We have what looks like an unimodal distribution, with a peak around 7-8 people. Note that in some cases the ICO team is composed just by the business and marketing people who developed the business idea – the developers will be hired only in the case of ICO success. When the software developers are part of the team, they typically account for a percentage between 20% and 50%. When a team is very large, this means that it include many advisors, who contribute suggestions but are not really involved in the ICO operations.

Fig. 5 shows the platforms used to deliver and manage the token or coin offer. As you can see, Ethereum is by far the most used platform. 193 ICOs do not declare their platform, and Waves is the second most popular platform, chosen by 67 ICOs. There are many other platforms or approaches that were used to deliver the tokens, but overall they cover only 54 cases. We also analyzed the smart contract standard used to manage the tokens. In 787 cases it is ERC-20 on Ethereum, in 581 cases it is not explicitly declared, and in 12 cases it is the new standard ERC223, which is an evolution of ERC-20. One ICO mentions the NEP5 standard, which is the equivalent of ERC-20 for NEO blockchain. The Ethereum standard ERC-20 for token management was developed in 2015. It defines a set of rules that a contract carried out with an Ethereum token has to implement [6]. The standard ensures the interoperability of the assets, making them more useful. These rules include, for example, how to transfer a token and how to access the data (name, symbol, supply, balance) of the token. Various implementations of ERC-20 written in Solidity language are freely available.

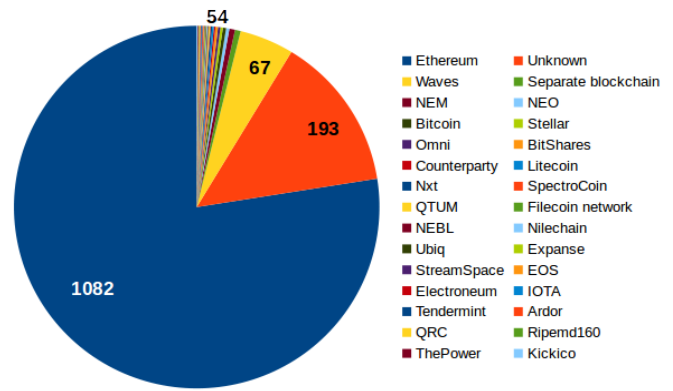

Fig. 5: The platforms used to manage the ICO token offer.

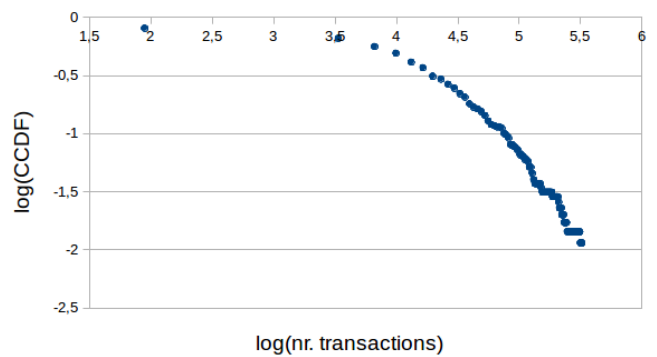

Fig. 6: CCDF of the transfers count of the considered ERC-20 tokens on the Ethereum blockchain.

From a software engineering perspective [10], it is worth noting the number of ICOs relying on Ethereum blockchain for the delivery and management of their tokens. It is the staggering number of 1082, steadily growing by the day and whose overall value overcomes 30-40 billion US$ at the present evaluation! Despite this load, Ethereum public blockchain looks performing quite well.

We analyzed the number of transfer transactions and of the token holders for all 355 tokens managed on Ethereum using ERC-20 standard, that also enables Web sites like ethplorer.io to easily gather and show relevant data. Table I shows the main statistics. As you can see, both data series have mean much higher than median, a high standard deviation and very large maximum values. This is a typical behavior of fat-tailed distributions. Consequently, we analyzed the distributions of these data, which are shown in Figs. 6 and 7 in the form of complementary cumulative distribution function (CCDF), in log-log format. Both distributions tend to follow a straight line in the right of the plot, which is the typical characteristic of power-law distributions.

### B. Multivariate Analysis of the factors influencing the success

We eliminated the ICOs still in progress, whose end date was in 2018, except for 5 ICOs that raised a significant amount of money in 2017, and were closed in advance. The ICOs ended within 2017 are 971. We also excluded the ICOs with no raised money according to icobench.com, and with no end date. We assumed they are ICOs still in progress, registered on the site, but whose end date is still to be determined – or even abandoned ICOs. The considered ICOs were thus reduced to 712. Among these ICO's tokens, only 215 are quoted on exchanges and their financial data are reported on coinmarketcap.com site. We were able to assess the third success criterion only for these tokens.

In order to perform the multivariate analysis we started including all the variables but those already excluded according to the described methodology. Those included in the full model are: rating, rtTeam, rtVision, rtProduct, rtProfile, country, platform, team size and finally category.

We briefly report the most significant values in Table II.

TABLE I. STATISTICS OF NUMBER OF TRANSFERS AND HOLDERS OF 355 ERC-20 ETHEREUM TOKENS.

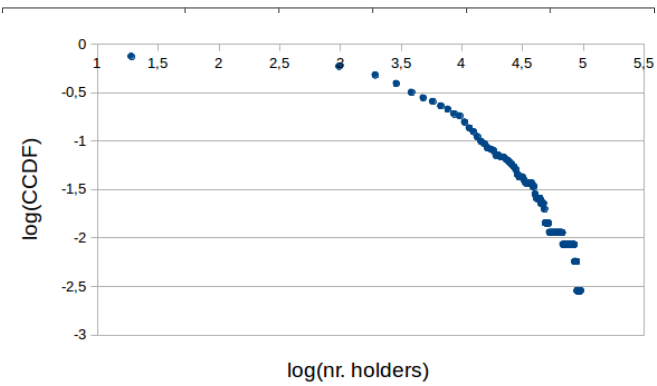

Fig. 7: CCDF of the holders count of the considered ERC-20 tokens.

| Data | mean | median | st. dev. | min | max |
|---|---|---|---|---|---|
| # of transfers | 46076 | 13186 | 132586 | 89 | 1311959 |
| # of holders | 15515 | 2872 | 76938 | 19 | 959205 |

TABLE II.

| Independent Variable | Coeff. | S.E. | Wald Z | Pr(>\|Z\|) |
|---|---|---|---|---|
| rating | 1.699 | 0.442 | 3.84 | 0.0001 |
| latform=Ethereum | -0.9867 | 0.2786 | -3.54 | 0.0004 |
| ountry=Slovenia | 1.6493 | 0.683 | 2.41 | 0.0158 |
| Categ...=Software | 1.1393 | 0.5329 | 2.14 | 0.0325 |
| ountry=USA | 0.7895 | 0.3816 | 2.07 | 0.0385 |
| rtProfile | -0.7512 | 0.3858 | -1.95 | 0.0515 |
| rtVision | -0.3291 | 0.2021 | -1.63 | 0.1033 |
| ountry=Israel | 1.2882 | 0.7912 | 1.63 | 0.1035 |
| ountry=China | 1.1639 | 0.7228 | 1.61 | 0.1073 |
| Categ...=CasinoGambling | 1.098 | 0.6861 | 1.6 | 0.1095 |
| rtProduct | 0.3379 | 0.2153 | 1.57 | 0.1167 |
| ateg...=Businessservices | -0.7462 | 0.5015 | -1.49 | 0.1367 |
|  |  |  |  |  |
| ountry=Singapore | 0.6228 | 0.4644 | 1.34 | 0.1799 |

| Independent Variable | Coeff. | S.E. | Wald Z | Pr(>|Z|) |
|---|---|---|---|---|
| rating | 1.699 | 0.442 | 3.84 | 0.0001 |
| ountry=UK | 0.5578 | 0.4475 | 1.25 | 0.2126 |
| latform=Waves | -0.5726 | 0.5039 | -1.14 | 0.2558 |

a. We only reported the most influencing variables

Table II reports the Logit model coefficients, their standard errors, the Wald normalized Z value and the relative p-value for all the independent variables in the model, sorted according to an increasing p-value, up to the case of the 'Wave' platform, which we retained since is the only one that can be compared with the most common platform 'Ethereum'.

The results show that the most significant variable is the 'rating' as reported by ICO-bench in a scale between 0 and 5. In paticular the relative p-value in the full model is 0.0001 and the coefficient value and the Z-Wald value are 1.6991 and 3.84 respectively. This means that the model identifies the variable relevant for influencing ICO's success according to the described criteria and that, in particular, a unit increase of the 'rating' carries a factor of about five in favor of the odd ratio, meaning that the odds are shifted of a consistent amount for each unitary increase of the 'rating' provided by icobench.com. The other way round, icobench.com rating system is a reliable indicator of the possible success of the ICO and, consequently, of the quality of the ICO project.

The other interesting variable related to icobench.com is the rtProfile with a 95% significance level (p-value 0.0515) and coefficient and Z-Wald of -0.7512 and 0.3858. This means that a unit increase of this index, which is in the range 0 to 5, raises the odd ratio of a factor of about 0.4. This indication is in agreement with the previous one, since the rtProfile is automatically assigned by a robot on the bases of a combination of values of the other four icobench.com indexes and is mainly influenced by the rating value, and coincides with it in the cases where the other indicators are missing.

The rtVision as well has some incidence on the success of the ICO, having a p-value around 0.1 and contributing to enhancing the odd ratio of a factor 0.7 for each unit increment.

Some interesting results concern the countries, the category and the platform. For the latter the topic case is the Ethereum platform. Data analysis shows that the Ethereum platform has high significance level (p-value 0.0004) and contributes to the odd ratio of a factor of about 0.4, on average, with respect to the other platforms.

It has to be noted that when considering data on platforms 'per-se' there are many spurious data, namely those where a given platform appears only once or in a very few cases.

In these particular cases, even if they are not at all statistically significant, the odd ratio is exceptionally high or low, meaning that the variable automatically means success or failure, given that they appear only twice or three times with always success or failure.

Not considering the spurious cases of platforms appearing one or very few times the only comparison can be made with Waves, another quite common platform, which, on the contrary, does not appear to provide a significant contribution to the ICO's success.

For what concern the countries the best ones where to start an ICO are Slovenia and USA. Good places are also, but to a less extent, Israel and China. In particular, Slovenia and USA have a good statistical significance, of about 0.016 and 0.038 respectively, and they contribute to the odd ratio in favor of success of about 5.2 and 2.2 respectively. The other countries have less statistical significance in determining the ICO's success.

Finally the category which positively contributes to the ICO's success is 'software', with a p-value of 0.0325 and a relative contribution to the odd ratio of 3.1. Other categories which in principle could be interesting are 'gambling' and 'business', but with a much lower statistical significance.

The analysis also shows that 'team size' does not seem to count for determining success or failure of the ICO project. Since we performed the analysis with the basic model using all available numerical data, we also checked the possible contribution of 'team size' to success or failure gathering into different categories the team's size, making different choices for the categories (small, medium, large tema's size or even 5 different categories), but also this analysis confirms that the variable does not count for ICO's success or failure.

IV. CONCLUSIONS

In this work we examined 1388 ICOs, from icobench.com Web site, also gathering information from other source. Financial information about the prices of the ICO tokens was obtained from coinmarketcap.com site, and transaction information coming from Ethereum blockchain was obtained from ethplorer.io site. An intial analysis gave insights on some key features of the ICOs, showing the countries they come from, the addressed business field, the team size and the software platforms used to manage the ICO tokens. Unsurprisingly, we found that most ICOs are managed on Ethereum blockchain, using ERC-20 standard. To this purpose, we also found that the distribution of token transfer transactions and token holders follow a fat-tailed distribution, resembling a power-law in the tail. This kind of distribution is very common in technological and financial data. Its meaning is that, though there are many tokens managed on the blockchain, only a few of them account for most of the workload applied to the blockchain.

Subsequently, we performed a multivariate analysis to assess the factors that can influence the success of an ICO. To this purpose, we divided the considered ICOs in two categories – successful and failed. The analysis showed that there are some factors that are correlated to an ICO success. They are the country of origin – it looks that ICOs coming from Slovenia and USA, and, to a less extent, Israel and China, are more prone to have success. Most other countries do not bear significance. The team size does not seem to be relevant to the success. A high overall rating on icobench.com site, on the other hand, looks quite correlated to the success of an ICO, though this looks mainly due to the robot's advice rather than to the human experts' advices. Finally, managing the ICO token on Ethereum blockchain looks another success factor.

Future work will regard gathering ICO data also from other sources, to double check their validity, and to perform a deeper analysis of token transactions on Ethereum blockchain, also to relate blockchain activity to price and volume information of the token.

ACKNOWLEDGMENT

We thank icobench.com administrators who granted us the permission to fully access their API interface.